\documentclass[prd,a4paper,showpacs,showkeys,preprint,byrevtex]{revtex4}
\usepackage{amsmath}
\usepackage{bm}
\begin{document}

\title{The baryonic Y-shape confining potential energy and its
approximants}

\author{Bernard \surname{Silvestre-Brac}}
\email[E-mail: ]{silvestre@lpsc.in2p3.fr}
\affiliation{Laboratoire de Physique Subatomique et de Cosmologie,
Av. des Martyrs 53, F-38026 Grenoble-Cedex, France}

\author{Claude \surname{Semay}}
\thanks{FNRS Research Associate}
\email[E-mail: ]{claude.semay@umh.ac.be}
\affiliation{Universit\'{e} de Mons-Hainaut, Place du Parc 20,
B-7000 Mons, Belgium}

\author{Ilya M. \surname{Narodetskii}}
\email[E-mail: ]{naro@heron.itep.ru}
\author{A. I. \surname{Veselov}}
\email[E-mail: ]{veselov@heron.itep.ru}
\affiliation{Institute of
Theoretical and Experimental Physics, B. Cheremushkinskaya 25,
117218 Moscow, Russia}
\date{\today}

\begin{abstract}
We discuss the validity of replacing the complicated three-body
confinement operator of the Y string junction type by three kinds
of approximation which are numerically much simpler to handle: a
one-body operator with the junction point at the centre of mass, a
two-body operator corresponding to half the perimeter of the
triangle formed by the three particles, and the average of both.
Two different approaches for testing the quality of the
approximations are proposed: a geometrical treatment based on the
comparison of the potential energy strengths for the various
inter quark distances, and a dynamical treatment based on the
comparison of the corresponding effective string tensions using a
hyperspherical approach. Both procedures give very similar
results. It is shown how to simulate the genuine string junction
operator by the approximations proposed above. Exact three-body
calculations are presented in order to compare quantitatively the
various approximations and to confirm our analysis.
\end{abstract}



\maketitle

\section{The Y string potential and possible approximations}
\label{sec1}

The QCD lattice calculations support the idea that the confining
potential energy in a three quark system can be, at best, simulated by
the so-called Y-shape potential \cite{taka01}. In this scheme,
the three pointlike quarks (located at the apices of a triangle $ABC$),
are connected by three flux tubes starting from a junction point ($I$)
in such a manner that it minimizes the sum of the distances to the three
quark positions. Note that effective QCD theories support also this idea
\cite{koma01}.

If the angle corresponding to an apex is greater than 120$^{\circ}$, the
junction point is precisely at this apex, whereas if all the angles are
less than 120$^{\circ}$, then the equilibrium junction position
coincides with the so-called Toricelli (or Fermat or Steiner) point; it
corresponds to a point for which the corresponding angles
$\widehat{AIB}$, $\widehat{BIC}$, and $\widehat{AIC}$ are all equal to
120$^{\circ}$.

Thus the genuine string potential in a baryon is defined by
\begin{eqnarray}
V_Y &=&\sigma (AB+AC) \quad \text{if} \quad
\widehat{A}>120^{\circ}\quad \text{(+ circular permutation on $B$
and $C$)},
\nonumber \\
V_Y &=&\sigma (IA+IB+IC) \quad \text{if} \quad \widehat{A},
\widehat{B}, \widehat{C} \leq 120^{\circ}, \label{VY}
\end{eqnarray}
where $\sigma$ is the string tension, in principle a constant of
the theory. This expression has been used in several works (see
for instance Refs.~\cite{artr75,dosc76,caps86,blas90}). The success of
simple quark models using a potential to simulate the confinement
without gluonic modes can be justified by the fact the lowest gluonic
excitation energy in the three quark system is found to be about 1~GeV
in the hadronic scale \cite{taka03}.

Efficient methods to deal with Y-shape interaction relies either on
Monte-Carlo algorithms (see for instance
Refs.~\cite{carl83,sart85}) or hyperspherical methods (see for
instance Ref.~\cite{fabr91}). However it is very difficult to
implement in a numerical code, essentially for two reasons: i) it
is a three-body operator which makes the equations very
complicated; ii) the presence of two different expressions
depending upon the configuration makes the integration domain not
evident to handle. This complexity explains why, for practical
calculations, the genuine Y-shape potential is often replaced by
other expressions that are considered as approximants.

The most popular one is to consider half the perimeter of the
triangle formed by the three quark coordinates and to define the
confining potential as
\begin{equation}
V_C=\frac{\sigma }{2}(AB+BC+CA),
\label{VC}
\end{equation}
which is also called the $\Delta$-shape potential. However it is shown
in Ref. \cite{taka01} that the three quark potential energy is better
represented by the Y-shape interaction than by the $\Delta$-shape one.
Nevertheless, potential (\ref{VC}) is a two-body operator, free from the
complication due to angles; hence it is much more simple and it is still
widely used in practice (see for instance
Refs.~\cite{bhad81,gloz98,brau02}).

A ${\bm \lambda_i} \cdot {\bm \lambda_j}$ colour dependence associated
with a two-body linear confinement does produce such a 1/2 factor.
Although this colour prescription is perfectly relevant in the case of
one-gluon exchange, there is no theoretical justification to apply it
for the confinement potential. This 1/2 factor is close to the value
0.53, predicted by lattice calculation \cite{taka01}, if one tries to
replace the Y-shape by a $\Delta$-shape.

It is very important to stress that both the genuine string
potential (\ref{VY}) and the two-body confining potential
(\ref{VC}) are of geometrical essence, depending only on the
position of the quarks and independent of their masses.

Another approximation, that was suggested, is to replace the true
junction  point $I$ by the centre of mass $G$ of the three quark
system   \cite{kerb00,naro02}. In this case the corresponding
confining operator is simply
\begin{equation}
V_G=\sigma (GA+GB+GC).
\label{VG}
\end{equation}
This approximation is particularly interesting since this
potential is a one-body operator free from angle complications;
hence its  numerical treatment is quite easy. In contrast to the
previous expressions, this approximation does depend on the system
via the centre of mass coordinate.

In this paper, assuming that $V_Y$ represents the true physics, we
want to study the quality of  the  approximations (\ref{VC}),
(\ref{VG})and their relevance. To this end, we propose two
approaches: a geometrical one and a dynamical one relying on the
hypercentral formalism. With this last technique, it is possible to
obtain directly an average value of the confining potential energy
depending only on one lenght parameter. Then we check their validity
with an exact three-body treatment based on a complete hyperspherical
treatment (beyond the hypercentral approximation). A simulation of
three-body potential by sums of two-body potentials was performed in
Ref.~\cite{fabr97}, but with a philosophy very different from the
one developed here.

In the next section, the geometrical approach is presented. In
Sec.~\ref{sec3}, the hyperspherical formalism is used to calculate
the value of the effective  string tension for a given system.
Sec.~\ref{sec4} is devoted to a three-body treatment of the
confining potential and the simulation of the genuine string
operator. Conclusions are drawn in the last section.

\section{Geometrical approach}
\label{sec2}

\subsection{Configuration of the system}

We are interested in the ratio of the potential energy for two
forms of confining interactions as function of the configuration
of the system. Since this ratio is scale independent and since the
dynamical constant $\sigma $ disappears in this ratio, it is
always possible to rescale the quark triangle putting  $AB$ = 1
and to deal only with the remaining apex of the triangle. Denoting
by $L$ the minimal distance from the junction point $I$, by $D$
the distance from the centre of mass $G$, and by $P$ half the
perimeter of the triangle, we will study the ratios:
$R_{Y/C}=V_{Y}/V_{C}=L/P$, $R_{G/C}=V_{G}/V_{C}=D/P$,
$R_{G/Y}=V_{G}/V_{Y}=D/L$.

In order to obtain analytical expressions (already complicated!)
for these ratios we restrict ourselves to a system with 2
identical quarks of mass $m$ located in $A$ and $B$ and a third
one of mass $M=xm$ located in $C$. The region $0<x<1$ corresponds
to $QQq$ systems with a light and two heavy quarks; the region
$x>1$ corresponds to $qqQ$ systems with a heavy and two light
quarks. The case $x=1$ corresponds to systems $qqq$ with three
identical masses.

Since $AB$ is fixed, the only freedom for the geometry of the
system is the position of the apex $C$, that can be defined by two
angles $a=\widehat{A}$ and $b=\widehat{B}$. Following the previous
remarks, one must calculate $L(a,b),P(a,b),D(a,b,x)$. In fact this
study is still too complicated. One can simplify it a lot noting
that
there is a symmetry versus the mediating line of $AB$, with the
consequence that $R(a,b)=R(b,a)$. This implies that an extremum of
those ratios, the only important thing for our consideration,
always lies in the mediating line and it is sufficient to restrict
the study to isosceles triangles. Thus, we only compute
$R_{Y/C}(a),R_{G/C}(a,x)$ and $R_{G/Y}(a,x)$.

In a quantum mechanical treatment of the confinement, the wave function
of course explores all the configurations for the triangle so that, may
be, the most important physical quantity to be computed is the average
of the previous ratios. One defines
\begin{equation}
R(x)= \frac{2}{\pi } \int_{0}^{\pi /2}R(a,x)\,da.
\label{Rx}
\end{equation}

\subsection{Various distances}

The calculation of the various distances results from geometrical
properties in a triangle and presents no difficulty. It is easy to show
that
\begin{itemize}
\item The genuine string distance is given by
\begin{subequations}
\label{la}
\begin{eqnarray}
L(a) &=&\frac{1}{\cos (a)}\quad \rm{if} \quad 0\leq a\leq \pi /6,
\label{la1} \\
L(a) &=&\frac{1}{2}(\tan (a)+\sqrt{3})\quad \rm{if}
\quad \pi /6<a<\pi /2. \label{la2}
\end{eqnarray}
\end{subequations}
\item Half the perimeter is equal to
\begin{equation}
P(a)=\frac{1+\cos (a)}{2\cos (a)}.
\label{pa}
\end{equation}
\item Centre of mass string distance is given by
\begin{equation}
D(a,x)=\frac{\tan (a)+\sqrt{x^{2}\tan ^{2}(a)+(2+x)^{2}}}{2+x}.
\label{da}
\end{equation}
\end{itemize}

\subsection{genuine potential to half perimeter}

With the expressions (\ref{la}) and (\ref{pa}), one has
$R_{Y/C}(a)=L(a)/P(a)$. This ratio is always larger than 1, meaning that
the genuine string junction potential is always more repulsive than the
sum of the two-body confining potentials. However the ratio is 1 for a
flat triangle or for an infinitely stretched triangle, and presents a
maximum, $2/\sqrt{3}\approx 1.155$, for an equilateral triangle.
This is in agreement with the result of Ref.~\cite{taka01}.

The average value (\ref{Rx}) in this case is equal to
\begin{equation}
R_{Y/C}=\frac{2}{\sqrt{3}}+\frac{2}{\pi }\left[ 1-\sqrt{3}+2\ln \left(
\frac{1+\sqrt{3}}{2}\right) \right] \approx 1.086.
\label{pe}
\end{equation}
The value $1/2 \, R_{Y/C} \approx 0.54$ must be compared with the
corresponding value 0.53 derived from QCD \cite{taka01}.
Thus the average error replacing the string tension operator by
the sum of the two-body confining potentials is of the order of
8\%; this approximation can be considered as a good one.

\subsection{Centre of mass junction approximation to half perimeter}

There does not exist special angle conditions in this case and,
following formulas (\ref{da}) and (\ref{pa}), the ratio is given
by  $R_{G/C}(a,x)= D(a,x)/P(a)$.

The maximum value is obtained for an infinitely stretched triangle
and for an infinite mass asymmetry. In this special case, the
ratio equals 2, but in the physical part of the domain, this ratio
is much closer to unity. A more reliable estimation results from
averaging following the procedure~(\ref{Rx}). The integral is
cumbersome but can be evaluated analytically using for
example the Mathematica package:
\begin{subequations}
\label{pgc}
\begin{multline}
R_{G/C}(x)=\frac{4}{\pi (2+x)}\Big\{ \ln (2)+x+2\sqrt{1+x}
\,\arctan \left(2\sqrt{1+x)}/x\right) \\
-(2+x)\,\text{E}(X)+x^{2}\,\text{K}(X)/(2+x)\Big\},
\label{pgc1}
\end{multline}
where
\begin{equation}
X =\frac{4(1+x)}{(2+x)^{2}},
\label{pgc2}
\end{equation}
\end{subequations}
and where $\text{K}(X)$ and $\text{E}(X)$ are the complete
elliptic integrals respectively of first and of second kinds
\cite{abra70}. The function $R_{G/C}(x)$ is presented in
Fig.~\ref{fig3}. It is always greater than 1, indicating that the
centre of mass string always overestimates the sum of the two-body
confining potentials. For $x=0$ ($QQq$ systems) the value is
1.168, for $x=1$ ($qqq$ systems) it is 1.136, and for $x=20$
($qqQ$ systems) it is 1.248. Curiously it passes through a minimum
1.132 for $x\approx 0.600$ corresponding to $ssu$ and $ssd$
systems. Replacing the two-body potential by the centre of mass
string induces an error of about 15-20\%.

\subsection{Centre of mass junction approximation to genuine potential}

In a similar way, we define the ratio $R_{G/Y}(a,x)=D(a,x)/L(a)$.
A maximum value 2 is obtained in very extreme
situation $a\rightarrow \pi /2$, $x\rightarrow \infty$,but, in general,
the values of expression this ratio are very close to 1.

The integration of expressions $R_{Y/C}(a)$ over the angle $a$ is
very cumbersome,
the final result looks like
\begin{subequations}
\label{rgy}
\begin{multline} R_{G/Y}(x)=\frac{4}{\pi (2+x)}\Bigg\{
\frac{1}{12}(\pi +3\sqrt{ 3} \ln (2))+\frac{2-\sqrt{3}}{4}
+\frac{2+x}{2}\,\text{E}(\pi /6,X)
\\
-\frac{1}{2} \sqrt{1+x+x^{2}}\ln (Y)+\sqrt{1+x}
\left(\frac{1}{4}\,\ln (T) +\frac{\sqrt{3}}{2} \arctan (U) \right)
\Bigg\} \label{rgya}
\end{multline}
where $X$ is given by Eq. (\ref{pgc2}) and the new quantities $Y$,
$T$, and $U$ are equal to
\begin{eqnarray}
Y &=&\frac{x(2\sqrt{1+x+x^{2}}-\sqrt{3}x)}{\sqrt{3}(1+x)+
\sqrt{(1+x+x^{2})(3+3x+x^{2})}}, \label{YZ1} \\
T &=&\frac{\sqrt{3+3x+x^{2}}-\sqrt{3(1+x)}}{\sqrt{3+3x+x^{2}}+
\sqrt{3(1+x)}}, \label{YZ2} \\
U &=& \frac{3 \sqrt{1+x}}{2 x + \sqrt{3+3x+x^2}}. \label{YZ3}
\end{eqnarray}
\end{subequations}
The behaviour of this ratio is shown on Fig.~\ref{fig5}. Some
remarkable values are $1.075$, $1.048$, $1.149$ for $x=0,1,20$.
The function $R_{G/Y}(x)$ passes through a minimum 1.043 for
$x\approx 0.585$, corresponding again to the $\Xi $ baryon. It is
very instructive to emphasize that for a large domain of masse
ratios $0<x<5$, the error introduced by replacing the Toricelli
point by the centre of mass is less than 10\%.

\subsection{Another approximation}

Replacing the genuine string junction potential by a sum of the
two-body confining potentials is a rather good approximation in
any case. Replacing it by a sum of one-body centre of mass string
potentials is even a better approximation for equal quark mass
systems and one light-two heavy quark ones. This approximation
becomes slightly worse (although not dramatically) for one
heavy-two light quark systems. In most cases, these approximations
are better than 10\%.

It is interesting to remark that the genuine string tension is
\emph{always comprised} between half perimeter and centre of mass
junction. This last property is obvious since the Toricelli point
is precisely the one which minimizes the sum of the distances; in
contrast the former property is by no means obvious. One can take
benefit of this remark
and to define a new ratio as
\begin{equation}
R_{M/Y}(x)=\frac{1}{2} \left( R_{C/Y} +R_{G/Y}(x) \right).
\label{rmy}
\end{equation}
In Eq. (\ref{rmy}), the function $R_{G/Y}(x)$ has been computed
before. The value of $R_{C/Y}$ can be computed analytically. The result
is
\begin{equation}
R_{C/Y}= \frac{1}{6\pi} \left( 3+ (1+\sqrt{3}) \pi - 3 \left[2\ln(
2) + \ln (\sqrt{3}-1) - 3\ln (\sqrt{3}+1) \right] \right) \approx
0.923. \label{rcy}
\end{equation}
The curve $R_{M/Y}(x)$ is plotted on Fig.~\ref{figsup}. One can
remark that the values of $R_{M/Y}(x)$ differ from unity by less
than 3\% for all relevant values of the $x$ parameter. So this
procedure to simulate the genuine string junction potential seems
preferable to the previous discussed ones.

\section{Hyperspherical approach}
\label{sec3}

\subsection{hyperspherical coordinates}
\label{hyper}

The hyperspherical formalism is an economical way to tackle the
three-body problem. We refer to specialized papers for technical aspects
(see for instance Ref.~\cite{fabr91}). Here we just recall what is
needed for our purpose.

Let us define a reference mass $m$ and introduce the dimensionless
quantities $\omega _{i}=m_{i}/m$, $\omega _{ij}=\omega _{i}+\omega
_{j}$ and $ \omega =\omega _{1}+\omega _{2}+\omega _{3}$. The
first thing to do is to introduce the Jacobi coordinates
\begin{equation}
{\bm \rho }_{ij}=\alpha _{ij}({\bm r}_{i}-{\bm r}_{j}),\quad{\bm
\lambda }_{ij}=\beta _{ij}\left( \frac{\omega _{i}{\bm r}_{i}+
\omega_{j}
{\bm r}_{j}}{\omega _{ij}}-{\bm r}_{k}\right),
\end{equation}
with
\begin{equation}
\alpha _{ij}=\sqrt{\frac{\omega _{i}\omega _{j}}{\omega
_{ij}}},\quad \beta _{ij}=\sqrt{\frac{\omega _{k}\omega
_{ij}}{\omega }},\quad \Omega =\alpha_{ij}\beta
_{ij}=\sqrt{\frac{\omega _{1}\omega _{2} \omega _{3}}{\omega}}.
\label{alpha_beta}
\end{equation}
The normalization quantities have been set in order to obtain nice
properties under particle permutations. From now on, we particularize to
the 1-2 pair and drop the 12 index everywhere, so that the Jacobi
coordinates for our problem are noted simply ${\bm \rho }$ (instead of
${\bm \rho } _{12}$) and ${\bm \lambda }$ (instead of
${\bm \lambda }_{12}$).

Each inter-distance ${\bm r}_{ij}={\bm r}_{i}-{\bm r}_{j}$ can be
expressed only in terms of ${\bm \rho }$ and ${\bm \lambda }$, so
that half the perimeter of the quark triangle is expressed as
$P({\bm \rho }^{2},{\bm \lambda } ^{2},{\bm \rho }\cdot {\bm
\lambda })$. The same is true for the positions relative to the
centre of mass ${\bm s}_{i}={\bm r}_{i}- {\bm R}_{cm}$, so that
the centre of mass string distance is $D({\bm \rho }^{2},{\bm
\lambda }^{2},{\bm \rho }\cdot {\bm \lambda })$. Finally, the same
property is valid for the genuine string distance $L( {\bm \rho
}^{2},{\bm \lambda }^{2},{\bm \rho }\cdot {\bm \lambda })$. Over
the six original variables defining the configuration, three have
disappeared corresponding to the three Euler angles giving the
orientation of the plane of the quarks in a fixed reference frame.
The confining potential is expressed in terms of two distances
$\rho $ and $\lambda$, and one angle $ \chi =(\widehat{\rho
},\widehat{\lambda})$ between ${\bm \rho }$ and ${\bm \lambda }$.
Instead of $\rho$ and $\lambda$, the hyperspherical formalism
introduces the hyperradius $R$ and the hyperangle $\theta $
through a polar transformation
\begin{equation}
\rho =R\sin \theta,\quad \lambda =R\cos \theta \quad \text{with} \quad
0 \le \theta \le \pi/2.
\end{equation}
The hyperradius is invariant under quark permutations
\begin{equation}
R=\sqrt{\rho ^{2}+\lambda ^{2}}=\sqrt{\rho _{23}^{2}+\lambda _{23}^{2}}=
\sqrt{\rho _{31}^{2}+\lambda _{31}^{2}}.
\end{equation}
The elementary volume element with the hyperspherical coordinates is
simply
\begin{equation}
dV=R^{5}dRd\Omega ^{(6)},\quad d\Omega ^{(6)}=\cos ^{2}\theta
\,\sin ^{2}\theta \,d\theta \,\sin \chi \,d\chi \,d\Omega ^{(3)},
\end{equation}
where $d\Omega ^{(6)}$ is the volume element on the hyperangles
and $d\Omega ^{(3)}$ the usual volume element on Euler angles. One
has obviously
\begin{equation}
\int d\Omega ^{(3)}=8\pi ^{2},\quad \int d\Omega ^{(6)}=\pi ^{3}.
\end{equation}

An important property of the hyperspherical formalism is that a
dominant part of the interaction comes from the hypercentral
approximation of the potential $V({\bm \rho},{\bm \lambda})$ which
is the average of the potential over the hyperangles
\begin{equation}
V(R)=\frac{1}{\pi ^{3}}\int V({\bm \rho },{\bm \lambda })\,d\Omega
^{(6)}.
\end{equation}
For potentials invariant under rotations, as the confining term we
are interesting in, the expression does not depend on Euler angles
so that, in practice, the hypercentral potential is simply
\begin{equation}
V(R)=\frac{8}{\pi }\int_{0}^{\pi /2}\cos ^{2}\theta \,\sin ^{2}\theta
\,d\theta \,\int_{0}^{\pi }V(R,\theta ,\chi )\,\sin \chi \,d\chi.
\label{vhpc}
\end{equation}
In the hypercentral approximation the complicated three-body problem
reduces to a differential equation including $V(R)$.
Within this approximation, we obtain directly an average value of the
potential confining energy depending only on the hyperradius. Such a
treatment is feasible for the Y-shape as well as its approximants.

In this paper we focus only on the genuine confining potential
$V_{Y}$ and its two approximations $V_{C}$ and $V_{G}$. In term of
the hyperradius $R$ and hyperangles $\theta$ and $\chi$, they are
defined by
\begin{subequations}
\label{vycg}
\begin{eqnarray}
V_{Y}(R,\theta ,\chi)&=&\sigma L(R,\theta ,\chi ), \label{vycg1}\\
V_{C}(R,\theta ,\chi)&=&\sigma P(R,\theta ,\chi ), \label{vycg2}\\
V_{G}(R,\theta ,\chi)&=&\sigma D(R,\theta,\chi ). \label{vycg3}
\end{eqnarray}
\end{subequations}
The hypercentral approximations of formulas (\ref{vycg}) resulting from
equation (\ref{vhpc}), are denoted here $V_{Y}(R)$, $V_{C}(R)$ and
$V_{G}(R)$.

The expressions for $P$ and $D$ are easy to obtain. For the moment,
let us quote the following formulas
\begin{equation}
D(R,\theta ,\chi )=\sum_{i<j}\frac{\beta _{ij}}{\omega _{k}}\lambda
_{ij}
\label{vg}
\end{equation}
and
\begin{equation}
P(R,\theta ,\chi )=\frac{1}{2}\sum_{i<j}\frac{\rho _{ij}}{\alpha _{ij}}.
\label{vc}
\end{equation}

The expression for $L$ is much more cumbersome. As was mentioned
above,  there are three special cases for which the angles $\Theta
_{i}=(\widehat{r_{ij}}, \widehat{r_{ik}})$ are greater than
120$^{\circ}$ ($-1<\cos \Theta _{i}<-1/2$); in this case
$L=r_{ij}+r_{ik}$. The ``normal case'' (all $\Theta _{i}$ less
than 120$^\circ$) is $L=r_{I1}+r_{I2}+r_{I3}$  and is less easily
obtained. Let us quote the final expression:
\begin{itemize}
\item First region
\begin{equation}
-1<\frac{\frac{\Omega }{\omega _{1}}\tan \theta +\cos \chi }{\sqrt{\frac
{
\Omega ^{2}}{\omega _{1}^{2}}\tan ^{2}\theta +2\frac{\Omega }{\omega
_{1}}
\tan \theta \cos \chi +1}}<-\frac{1}{2}
\label{z1},
\end{equation}
\begin{equation}
L_{1}(R,\theta ,\chi )=R\cos \theta \left[ \frac{\tan \theta }{\alpha }+
\frac{1}{\beta }\sqrt{\frac{\Omega ^{2}}{\omega _{1}^{2}}\tan ^{2}\theta
+2
\frac{\Omega }{\omega _{1}}\tan \theta \cos \chi +1}\right].
\label{l1}
\end{equation}
\item Second region
\begin{equation}
-1<\frac{\frac{\Omega }{\omega _{2}}\tan \theta -\cos \chi }{\sqrt{\frac
{
\Omega ^{2}}{\omega _{2}^{2}}\tan ^{2}\theta -2\frac{\Omega }{\omega
_{2}}
\tan \theta \cos \chi +1}}<-\frac{1}{2},
\label{z2}
\end{equation}
\begin{equation}
L_{2}(R,\theta ,\chi )=R\cos \theta \left[ \frac{\tan \theta }{\alpha }+
\frac{1}{\beta }\sqrt{\frac{\Omega ^{2}}{\omega _{2}^{2}}\tan ^{2}\theta
-2
\frac{\Omega }{\omega _{2}}\tan \theta \cos \chi +1}\right].
\label{l2}
\end{equation}
\item Third region
\begin{equation}
-1<\frac{1+\Omega \left( \frac{1}{\omega _{1}}-\frac{1}{\omega _{2}}
\right)
\tan \theta \cos \chi -\frac{\Omega ^{2}}{\omega _{1}\omega _{2}}\tan
^{2}\theta }{\sqrt{1+2\frac{\Omega }{\omega _{1}}\tan \theta \cos \chi +
\frac{\Omega ^{2}}{\omega _{1}^{2}}\tan ^{2}\theta }\sqrt{1-2\frac{
\Omega }
{\omega _{2}}\tan \theta \cos \chi +\frac{\Omega ^{2}}{\omega _{2}^{2}}
\tan
^{2}\theta }} <-\frac{1}{2},
\label{z3}
\end{equation}
\begin{multline}
L_{3}(R,\theta ,\chi ) =\frac{R\cos \theta }{\beta }
\Bigg[ \sqrt{1+2\frac{\Omega }{\omega _{1}}\tan \theta \cos \chi +\frac
{\Omega ^{2}}{\omega _{1}^{2}}\tan ^{2}\theta } \\
+ \sqrt{1-2\frac{\Omega }{\omega _{2}}\tan \theta \cos \chi
+\frac{\Omega ^{2}}{\omega _{2}^{2}}\tan^{2}\theta }\Bigg].
\label{l3}
\end{multline}
\item Fourth region or ``normal region"
\begin{equation}
-1/2 < \cos \Theta_i < 1 \quad \forall\, i,
\label{zy}
\end{equation}
\begin{multline}
\label{ly}
L_4(R,\theta ,\chi ) = R\cos \theta  \\ \times
\sqrt{\dfrac{\omega _{1}^{3}-
\omega_{2}^{3}}{\omega _{1}\omega _{2}(\omega_{1}^{2}-\omega _{2}^{2})}
\tan ^{2}\theta +\left[ \dfrac{\omega _{2}-\omega
_{1}}{\omega _{12}}\cos \chi +\sqrt{3}\sin \chi \right] \dfrac{\tan
\theta}{\Omega }+\dfrac{1}{\beta^2}}.
\end{multline}
\end{itemize}
Since each distance is proportional to the hyperradius, all the
confining potentials under consideration have the form ($I=C,G,Y$)
\begin{equation}
V_{I}(R)=b_{I}(\omega _{1},\omega _{2},\omega _{3})R.
\end{equation}

One sees that in the hyperspherical formalism the confining
potential remains linear, with the string tension which is no
longer $\sigma$ but a modified value $b$ which depends on the
system. This dependence is not of geometrical essence, as in the
previous Section for $V_{G}$, but of dynamical character that
comes from the choice of the Jacobi coordinates. The main effort
of this section is devoted to the calculation of this new
``string'' tension $b$.

To stick to the previous Section and also to get analytical
results, we restrict from now on to the case of two identical
particles (particles 1 and 2 with the reference mass
$m=m_{1}=m_{2}$ and a particle 3  with mass $m_{3}=xm$). In this
special case, the coefficients $\alpha$ and $\beta$  in Eqs.
(\ref{alpha_beta}) are
\begin{equation}
\alpha =\frac{1}{\sqrt{2}},\quad \beta
=\sqrt{\frac{2x}{2+x}},\quad \Omega =\alpha \beta=
\sqrt{\frac{x}{2+x}}. \label{special}
\end{equation}
Let us study separately each approximation.

\subsection{Centre of mass junction}

Among the three contributions to the potential (\ref{vg}), one is
particularly simple. It corresponds to $s_{3}=(\beta R\cos \theta
)/\omega _{3}$. The averaging is trivial and gives the value
$32\beta R/(15\pi \omega _{3})$. The calculation of $s_{1}$ is
longer but presents no special difficulty. A trick to fasten the
answer is to remark that $R$ is an invariant under quark
permutations. Consequently, the averaging can be performed with
the hyperspherical angles $\theta _{23}$ and $\chi _{23}$ as well,
so that the result follows immediately from the previous one
giving the contribution $32\beta _{23}R/(15\pi \omega _{1})$.
Similarly the contribution due to $s_{2}$ is $ 32\beta
_{31}R/(15\pi \omega _{2})$. Replacing the quantities $\beta_{ij}
,\omega_k$ by their values (\ref{special}) gives the final result
for the effective string tension in this approximation
\begin{equation}
b_{G}(x)=\frac{32}{15\pi }\sigma \left[ \sqrt{\frac{2}{x(2+x)}}+2\sqrt{
\frac{1+x}{2+x}}\right].
\label{bg}
\end{equation}

\subsection{Half the perimeter}

The procedure is essentially the same using now potential
(\ref{vc}). The contribution due to $r_{12}=\rho/\alpha $ is very
easy to obtain; the result is $32R/(15\pi \alpha )$. Switching to
the permuted hyperangles gives immediately the contributions due
to $r_{23}$ and $r_{31}$, namely $32R/(15\pi \alpha _{23})$ and
$32R/(15\pi \alpha _{31})$ respectively. Replacing the quantities
$\alpha_{ij}$ by their values (\ref{special}) gives the final
result for the effective string tension in this approximation
\begin{equation}
b_{C}(x)=\frac{32}{15\pi }\sigma \left[ \frac{1}{\sqrt{2}}+\sqrt{\frac{1
+x}{x}}\right].
\label{bc}
\end{equation}

\subsection{Genuine string junction}

Because of the special cases, several zones of the plane
($\theta,\chi$) must be isolated corresponding to the conditions
((\ref{z1}), (\ref{z2}), (\ref{z3})), where the integrand has a
special form ((\ref{l1}), (\ref{l2}), (\ref{l3})). These various
zones are separated from the ``normal zone" by a line which
corresponds to the limit $\cos \Theta_i =-1/2$. It is  convenient
to take $\theta$ as abscissa and $\chi$ as ordinate in the plane.

The condition (\ref{z1}) writes explicitly
\begin{equation}
\chi _{1}(\theta )=\arccos \left[ \frac{-3\Omega \tan \theta -\sqrt
{4-3\Omega ^{2}\tan ^{2}\theta }}{4}\right].
\label{cdt1}
\end{equation}
This curve starts with the value $2\pi /3$ for $\theta =0$ and ends up
to $\pi $ for $\theta =\theta _{0}=\arctan \left( 1/\Omega \right) $. In
the first region, above $\chi _{1}$, the expression $L=L_{1}$ (\ref{l1})
applies.

The condition (\ref{z2}) writes explicitly
\begin{equation}
\chi _{2}(\theta )=\arccos \left[ \frac{3\Omega \tan \theta +
\sqrt{4-3\Omega^{2}\tan ^{2}\theta }}{4}\right] =\pi -\chi _{1}(\theta
).
\label{cdt2}
\end{equation}
This curve starts with the value $\pi /3$ for $\theta =0$ and ends up to
$0$ for $\theta =\theta _{0}$. In the second region, below $\chi _{2}$,
the expression $L=L_{2}$ (\ref{l2}) applies.

Finally a third region is found from condition (\ref{z3}), which
writes explicitly in terms of two functions
\begin{equation}
\chi _{3}^{-}(\theta )=\arccos \left[ \frac{\sqrt{(3\Omega ^{2}\tan
^{2}\theta -1)(3-\Omega ^{2}\tan ^{2}\theta )}}{2\Omega \tan \theta }
\right],\quad
\chi _{3}^{+}(\theta )=\pi -\chi _{3}^{-}(\theta ).
\label{cdt3}
\end{equation}
These curves start with the values 0 and $\pi $ for
$\theta =\theta _{0}$ and join to the common value $\pi /2$ for
$\theta =\theta _{1}=\arctan (\sqrt{3}/\Omega )$. In the third
region, at the right of $\chi _{3}$, the expression $L=L_{3}$ (\ref{l3})
applies.

In the rest of the plane, the ``normal'' expression $L_4$
(\ref{ly}) is valid. The situation is summarized on
Fig.~\ref{fig6}.

To obtain $b_Y(x)$ we integrate the function $L(R,\theta,\chi)$ in
the plane $\theta,\chi$.  The result is given  as the sum of seven
contributions
\begin{equation}
I=\frac{1}{\pi ^{3}}\int L(R,\theta ,\chi )\,d\Omega
^{(6)}=\frac{8\alpha R}{\pi}\left( I_{1}+I_{2}+I_{3}+I_{4}+I_{5}+I_{6}+
I_{7}\right).
\end{equation}
We were not able to obtain an analytical formula for $I$.  The
best that we can do is to transform the double integral into a
single integral noting  that the integrals over $\chi $ can be
calculated  analytically. Instead of integration over $\theta $,
it is much simpler to perform the change of variable $Z=\Omega
\tan \theta $. With these options, one has explicitly
\begin{subequations}
\label{int}
\begin{eqnarray}
I_{1} &=&\Omega ^{3}\int_{0}^{1}\frac{Z^{2}\,dZ}{(Z^{2}+\Omega ^{2})^{7/
2}}
\int_{0}^{\chi _{2}(Z)}\left[ 2Z+\sqrt{1+Z^{2}-2Z\cos \chi }\right] \sin
\chi \,d\chi, \label{int1} \\
I_{2} &=&\Omega ^{3}\int_{0}^{1}\frac{Z^{2}\,dZ}{(Z^{2}+\Omega ^{2})^{7/
2}}
\int_{\chi _{2}(Z)}^{\chi _{1}(Z)}\left[ \sqrt{1+3Z^{2}+2\sqrt{3}Z\sin
\chi }
\right] \sin \chi \,d\chi, \label{int2} \\
I_{3} &=&\Omega ^{3}\int_{0}^{1}\frac{Z^{2}\,dZ}{(Z^{2}+\Omega ^{2})^{7/
2}}
\int_{\chi _{1}(Z)}^{\pi }\left[ 2Z+\sqrt{1+Z^{2}+2Z\cos \chi }\right]
\sin
\chi \,d\chi, \label{int3} \\
I_{4} &=&\Omega ^{3}\int_{1}^{\sqrt{3}}\frac{Z^{2}\,dZ}{(Z^{2}+\Omega
^{2})^{7/2}} \nonumber \\
&&\quad\times \int_{0}^{\chi _{3}^{-}(Z)}\left[ \sqrt{1+Z^{2}+2Z\cos
\chi }+
\sqrt{1+Z^{2}-2Z\cos \chi }\right] \sin \chi \,d\chi, \label{int4} \\
I_{5} &=&\Omega ^{3}\int_{1}^{\sqrt{3}}\frac{Z^{2}\,dZ}{(Z^{2}+\Omega
^{2})^{7/2}}\int_{\chi _{3}^{-}(Z)}^{\chi _{3}^{+}(Z)}\left[ \sqrt{1+3Z^
{2}+2
\sqrt{3}Z\sin \chi }\right] \sin \chi \,d\chi, \label{int5} \\
I_{6} &=&\Omega ^{3}\int_{1}^{\sqrt{3}}\frac{Z^{2}\,dZ}{(Z^{2}+\Omega
^{2})^{7/2}} \nonumber \\
&&\quad\times\int_{\chi _{3}^{+}(Z)}^{\pi }\left[ \sqrt{1+Z^{2}+2Z\cos
\chi }+
\sqrt{1+Z^{2}-2Z\cos \chi }\right] \sin \chi \,d\chi, \label{int6} \\
I_{7} &=&\Omega ^{3}\int_{\sqrt{3}}^{\infty }\frac{Z^{2}\,dZ}{(Z^{2}+
\Omega
^{2})^{7/2}} \nonumber \\
&&\quad\times\int_{0}^{\pi }\left[ \sqrt{1+Z^{2}+2Z\cos \chi }+\sqrt{1+Z
^{2}-2Z\cos
\chi }\right] \sin \chi \,d\chi. \label{int7}
\end{eqnarray}
\end{subequations}
Due to the symmetries of the problem, it is easy to show that
$I_{1}=I_{3}$ and $I_{4}=I_{6}$. The integrals $I_{1}$, $I_{4}$, $I_{7}$
can be calculated analytically, but not $I_{2}$ and $I_{5}$.

The calculation is not trivial, but some help is provided with the
Mathematica package. One obtains
\begin{subequations}
\label{by}
\begin{eqnarray}
b_{Y}(x) &=&\frac{32\sigma }{15\pi }\sqrt{2}\left\{ \frac{28+39\Omega
^{2}+18\Omega ^{4}}{4(4+3\Omega ^{2})^{2}}\right. \nonumber \\
&&+\frac{120+368\Omega ^{2}+612\Omega ^{4}+641\Omega ^{6}+377\Omega
^{8}+105\Omega ^{10}+9\Omega ^{12}}{2\Omega (4+3\Omega ^{2})^{2}(3+
\Omega
^{2})^{2}(1+\Omega ^{2})^{3/2}} \nonumber \\
&&+\frac{5\Omega ^{3}}{4\sqrt{3}}\int_{0}^{1}\frac{dZ\,Z(1+\sqrt{3}Z)}
{(Z^{2}+\Omega ^{2})^{7/2}}  \nonumber \\
&&\qquad\times\left[ (1+3Z^{2})\text{E}\left(\frac{\pi }{4}-\frac{\chi
_{2}}{2},u\right)-(1-
\sqrt{3}Z)^{2}\text{F}\left(\frac{\pi }{4}-\frac{\chi _{2}}{2},u\right)
\right] \nonumber \\
&&+\frac{5\Omega ^{3}}{4\sqrt{3}}\int_{1}^{\sqrt{3}}\frac{dZ\,Z(1+\sqrt{
3}Z)}{(Z^{2}+\Omega ^{2})^{7/2}} \nonumber \\
&&\qquad\times\left. \left[ (1+3Z^{2})\text{E}\left(\frac{\pi }{4}-\frac
{\chi_{3}^{-}}{2}
,u\right)-(1-\sqrt{3}Z)^{2}\text{F}\left(\frac{\pi }{4}-\frac{\chi _{3}^
{-}}{2},u\right)
\right] \right\}
\label{by2}
\end{eqnarray}
with
\begin{eqnarray}
\Omega (x)
&=&\sqrt{\frac{x}{2+x}},\quad u(Z)=\frac{4\sqrt{3}Z}{(1+\sqrt{3}
Z)^{2}},\nonumber \\
\chi _{2}(Z) &=&\arccos \left[ \frac{3Z+\sqrt{4-3Z^{2}}}{4}\right]
,\quad \chi_{3}^{-}(Z)=\arcsin \left[
\frac{\sqrt{3}(Z^{2}-1)}{2Z}\right],
\label{omuchi}
\end{eqnarray}
and F$(\phi,m)$ and E$(\phi,m)$ are respectively the
elliptic integrals of first and second kinds \cite{abra70}.
\end{subequations}

\subsection{Comparison of the effective string tensions}
\label{compst}

It is interesting to compare the string tensions $b_{G}(x)$ (\ref{bg}),
$b_{C}(x)$ (\ref{bc}), $b_{Y}(x)$ (\ref{by}) for the three expressions
of the confining operator. Contrary to the previous geometrical approach
they all depend on the system; as in the preceding section we introduce
the ratios of the string tensions $r_{Y/C}(x)=b_Y(x)/b_C(x)$,
$r_{G/C}(x)=b_G(x)/b_C(x)$, $r_{G/Y}(x)=b_G(x)/b_Y(x)$.

\subsubsection{Genuine to perimeter}

The ratio $r_{Y/C}(x)$ is presented on Fig.~\ref{fig7}. It is always
greater than one, indicating that the genuine tension is more repulsive
than the two-body approximation. It starts from 1 for $x=0$, increases
to a maximum 1.099 for $x\approx 1$ and then decreases slowly around
1.085
for large values of $x$. Those values are very close to the constant
value 1.086 obtained in the geometrical approach.

\subsubsection{Centre of mass to perimeter}

The ratio $r_{G/C}(x)$ is plotted on Fig.~\ref{fig3}. It starts from 1
for $x=0$, increases rapidly up to 1.15 for $x\approx 0.5$ and then
presents a plateau at this value around $x\approx 1$; it tends
asymptotically to $2\sqrt{2}/(1+\sqrt{2})\approx 1.1715$ for an infinite
value of $x$. Although the forms of the curves for the hyperspherical
formalism and the geometrical approach are not identical, the values for
the corresponding ratios are rather close.

\subsubsection{Centre of mass to genuine}

The ratio $r_{G/Y}(x)$ is shown on Fig.~\ref{fig5}. It starts from
1 for $x=0$, raises to a local maximum 1.0556 for $x\approx
0.195$, then passes through a local minimum 1.051 for $x\approx 1$ and
raises very slowly to 1.08 for large values of $x$. Here again the
values are in nice agreement with the geometrical approach, and
are even closer to the genuine case.

\subsubsection{Another approximation}

In the hyperspherical formalism, we find the same features
concerning the approximations than in the geometrical approach:
for a given value of the string constant $\sigma$, the centre of
mass junction potential is more repulsive than the potential with
the genuine junction which is in turn more repulsive than half the
perimeter potential. These two approximations can be considered as
quite reasonable, differing by no more than 10\% as compared to
the genuine string junction one.

As in the case of the geometrical approach, one remarks that
$b_Y(x)$ is always comprised between $b_C(x)$ and $b_G(x)$. Thus
it is tempting to define a new approximation $b_M(x)$ and a new
ratio $r_{M/Y}(x)$ similar to that of Eq.~(\ref{rmy}) by
\begin{equation}
r_{M/Y}(x)=\frac{b_M(x)}{b_Y(x)} =
\frac{(b_C(x)+b_G(x))/2}{b_Y(x)} = \frac{1}{2} \left( r_{C/Y}(x)
+r_{G/Y}(x) \right). \label{rmy2}
\end{equation}
The curve $r_{M/Y}(x)$ is plotted on Fig.~\ref{figsup}. One can
remark from this Figure that $r_{M/Y}(x)$ differ from unity by less
than 2\% for all values of the parameter $x$. Like in the geometrical
case, using this prescription to simulate the genuine string
junction potential seems more preferable to the ones discussed
previously.

\section{Three-body calculations}
\label{sec4}

\subsection{Strategy}

For a system composed of two identical particles of mass $m$ and a
third particle of mass $M=xm$, the three-body equation in  the
hypercentral  approximation to the hyperspherical formalism, is
very simple; it looks like
\begin{equation}
[K(x,R)+b(x)R]\Psi(R)=E \Psi(R).
\label{tbhyp}
\end{equation}
In this expression $E$ is the energy eigenvalue, and $K(x,R)$
symbolizes the differential operator corresponding to the kinetic
energy plus all the hypercentral potentials except the confining
hypercentral interaction which is explicitly written as $b(x)R$,
as it was proved in the preceding section. Up to now, we
considered three types of confining potentials $V_I$, labelled by
an index $I$ ($I=Y$ for the genuine three-body confinement, $I=C$
for the two-body half perimeter confinement and $I=G$ for the
one-body centre of mass junction). The important point is that,
for these three possibilities, the operator $K(x,R)$ is {\em the
same}. For a given string tension  $\sigma$, the only difference
lies in a different value of the constant $b(x)$ appearing in
Eq.~(\ref{tbhyp}). For a genuine string tension,  the value
$b(x)=b_Y(x)$ given by Eq. (\ref{by}), while for centre of mass
junction and half perimeter confinement one must employ $b_G(x)$
and $b_C(x)$ of formulas (\ref{bg}) and (\ref{bc}) respectively.
The numerical results of the various possibilities are of course
different although they should be close, within less than 10\%.
This property was the conclusion of our two previous sections.

At the light of the behaviour of the curves $R_{M/Y}(x)$ and
$r_{M/Y}(x)$, it is natural to try to simulate the potential $V_Y$
by defining a new confining potential
\begin{equation}
V_M =\frac{1}{2} \left( V_C + V_G \right),
\label{VM}
\end{equation}
with $V_C$ and $V_G$ given respectively by formulas~(\ref{VC}) and
(\ref{VG}). Let us emphasize that the unique above definition
agrees with both the geometrical approach for the ratio
$R_{M/Y}=V_M/V_Y$ and the hyperspherical approach using $b_M(x)$.

As the ratios $R_{M/Y}(x)$ and $r_{M/Y}(x)$ are very close to 1,
one can hope that the $V_M$ and the $V_Y$ will give similar
spectra with the same string tension. The situation is not so
favourable for potentials $V_C$ and $V_G$ since the corresponding
ratios $R$ and $r$ can differ from 1 by more than 10\%.
Nevertheless, we can try to simulate results obtained from the
Y-shape potential by using a renormalized value of the string
tension in potentials $V_C$ and $V_G$.

Indeed, let us suppose that we perform an hyperspherical
calculation for $I$ approximation with a string constant
$\sigma_I$. Now let us perform an hyperspherical calculation for
$J$ approximation not with an identical string constant
$\sigma_J=\sigma_I$, but with a string constant modified in the
following way $\sigma_J(x)=\sigma_I\, r_{I/J}(x)$
($\sigma_J(x)=\sigma_I\, b_I(x)/b_J(x)$). Obviously, one recovers
the original equation and thus the original results. In other
words, one can simulate the results of a treatment based on an
approximation $I$, by performing a treatment based on an
approximation $J$, provided we change the string constant in a
consistent way. This conclusion is perfectly exact for the
hypercentral approximation. This does not mean that it must remain
exact if we perform a more sophisticated three-body treatment. The
quality of this simulation, as well as the results coming from
$V_M$, are the subjects of this section.

Our numerical algorithm to solve the three-body problem with
potentials $V_C$, $V_G$ and $V_M$ is based on an expansion of the
wave function in terms of harmonic oscillator functions with
different sizes \cite{nunb77}. It was checked with other methods
and was proved to give results of good accuracy if the expansion
is pushed sufficiently far (let say up to 16-20 quanta). Moreover
it can deal easily with a relativistic kinetic energy operator.
The detail of technical aspects is not the subject of this paper
and can be found elsewhere \cite{silv01}. For the present purpose,
it is enough to say that we are able to solve in a very fast and
precise way a three-body calculation either with a
non-relativistic or relativistic expression for the kinetic energy
operator.

One-body and two-body operators are easy to implement, but
three-body operators are much more complicated to handle,
specially if one must distinguish several integration domains, as
it is the case for the Y-shape potential. In this paper, the
three-body problem with potentials $V_Y$ is solved by the
hyperspherical method without the limitation of the hypercentral
approximation \cite{fabr91}. At the present stage, only S-wave
states can be computed with good accuracy. This is why the
simulation of a three-body operator either with a two-body, a
one-body operator, or a mixing of both is important.

We consider a system composed
of 3 quarks of type $n$ (for $u$ or $d$), $s$, $c$, $b$. In this
paper, we are only interested in the possibility to simulate the
genuine confinement by simpler potentials and, for that purpose,
it is enough to restrict the interaction to the confining term
only. Forgetting about Coulomb, hyperfine and constant potentials,
our results cannot be compared to physical systems. But the
comparison of the various simulations between themselves are very
instructive. But just to have some connections with the real
systems, we put arbitrarily the masses of the quarks at the physical
values (in GeV) \cite{bhad81}: $m_n=0.330$, $m_s=0.550$, $m_c=1.850$,
$m_b=5.200$. The only potential taken into account is the linear
confining potential as defined in the first section. To see the
sensitivity of the results versus the kinematics, we perform two types
of calculation: one based on a non-relativistic expression
(Schr\"{o}dinger equation), and one based on a relativistic
expression (spinless Salpeter equation). To test also the sensitivity to
excited states, we performed the calculations not only for ground
states ($L=0$ and $N=1$), but also for the first radial excited state (
$L=0$ and
$N=2$), and when it is possible the first orbital excited state ($L=1$
and $N=1$).

\subsection{Comparison centre of mass to perimeter}


This section does not deal with the Y shape potential, but nevertheless
it is important because we compare here two treatments that we are able
to handle easily, in any physical situation, and with a good accuracy.
The possibility to simulate one treatment by the other is very
instructive
and can be tested carefully, so that firm conclusions can be drawn.

For a number of systems, exploring a large domain of the $x$ parameter,
we calculate first the baryon binding energies obtained with the centre
of mass junction ($G$ approximation) and with the half perimeter
confinement ($C$ approximation) using the same value of the string
tension $\sigma=0.2$ GeV$^2$. This value is close to the accepted value
coming from lattice calculations \cite{taka01}. Then we keep this value
of $\sigma$, do the calculations for the $G$ approximation using a
modified value of the string tension based on the arguments of the
geometrical approach $\sigma_R(x)=\sigma\, R_{C/G}(x)$ and on the
arguments of the hyperspherical approach $\sigma_r(x)$=
$\sigma\, r_{C/G}(x)$, and compare with the $C$ calculation with the
string tension at the value $\sigma$. We also perform the reciprocal
calculations. Our quantitative results are presented in Table~\ref{tab1}
for a nonrelativistic kinematics.

Note that the definition of the ratio $r_{I/J}$ (see
sec.~\ref{compst}) implies that $r_{J/I}=1/r_{I/J}$. This property
is not exact for the ratio $R_{I/J}$. However, the value
$1/R_{C/Y}\approx 1.083$ (coming from formula~(\ref{rcy})) must be
compared with $R_{Y/C}\approx 1.086$ (formula~(\ref{pe})). These
values are very close and this gives us confidence to use
$R_{J/I}\approx 1/R_{I/J}$ for all cases.

If we compare column (1) and column (4) from Table~\ref{tab1}, we can
see that, with the same string tension, masses obtained by potentials
$V_C$ or $V_G$ are rather different. The binding energy can differ by
about 200 MeV or more. So it is relevant to answer the question: ``Is it
possible to simulate each potential by the other, simply in adjusting
the value of the string tension?".

If we look at columns (1) and (2) for systems characterized by $x=1$, we
can see that all masses are identical (actually, they differ at the 6th
digit, which is at the limit of the accuracy of our calculation method).
It is then possible to simulate perfectly the potential $V_C$ with a
string tension $\sigma$ by the potential $V_G$ with the string tension
$\sigma\, r_{C/G}$. This property, which is exact at the hypercentral
approximation, is also verified for the full calculation for symmetrical
systems. The situation is less favourable for systems with $x \ne 1$. In
these cases, we can remark that the masses of column (2) are always
greater than the masses of column (1). For $snn$ and $nss$ baryons, the
agreement is good, especially for the ground state. But for very
asymmetrical systems such as $bnn$ and $nbb$ baryons, there can exist
greater mass differences, up to about 100 MeV. Let us remark that, for
the ground state, the agreement is still reasonable (better than 2\%).

If we compare now data from columns (1) and (3), we can see that, in
general, the potential $V_C$ is not so well simulated by the potential
$V_G$ using $R_{C/G}$ instead of $r_{C/G}$. In particular, masses from
these two columns are not identical for systems with $x=1$ (since an
exact simulation is impossible with the geometrical approach, we
consider that using $1/R_{C/G}$ instead of $R_{G/C}$ in column (6) do
not spoil our conclusions). In most cases, masses of columns (3) are
greater than masses of column (2). But for very asymmetrical systems, a
peculiar mass in column (3) can be closer to the one in column (1) than
the corresponding mass in column (2).

We retrieve the same features in comparing data from columns (4)
with data from columns (5) and (6). But one can remark that the
masses of column (5) are always smaller of the masses of column
(4), when they are not identical. Moreover the masses of column
(6) are generally smaller than those of column (5). This situation
is the opposite of the one for columns (1), (2) and (3)

Some calculations have also been performed with a relativistic
kinematics, but the same conclusions can be drawn. We have just remarked
that the agreement between masses is slightly less good than in the
nonrelativistic case. To be complete, let us mention that, with this
kinematics, for systems
with $x=1$, the differences between masses computed with $V_C$ ($V_G$)
and those computed with $V_G$ ($V_C$) and the string tension multiplied
by $r_{C/G}$ ($r_{G/C}$) appear at the 5th digit, which can be
considered as relevant for the accuracy of our method.

The simulation based on the renormalized string constant originated from
the hyperspherical formalism gives generally good results, better than
those coming from  the geometrical approach. Nevertheless, for very
asymmetrical systems, better results can be obtained with this last
method.

It is important to emphasize that using renormalized string tensions
provides, in any case, much better results (6\% in the worse case)
than keeping a single value of $\sigma$ (17\% in the worse case,
10\% in the best). This study gives a strong confidence to use also
a renormalized $\sigma$ to simulate the genuine string junction.

\subsection{Simulation of the genuine junction}

We consider the genuine confining three-body potential with a
constant string tension $\sigma$ and its various approximations.
Our quantitative results are presented in Table~\ref{tab2} for a
nonrelativistic kinematics.

In this paper, the spectra of the potential $V_Y$ is obtained
by a hyperspherical formalism containing grand momenta $K=0,2,4$.
For the moment, only binding energies of S-wave states have been
computed. They are reported in column (1) and are used as a
reference to test the quality of the various approximations. All
other values have been computed with a harmonic oscillator basis
up to 20 quanta. The accuracy of all binding energies is better
than 1\%.

In column (2), binding energies of the potential $V_M$ are
presented. They are obtained with a value of string tension which
is the same that the one used for $V_Y$. One can see that the numbers of
columns (1) and (2) differ generally by less than 20 MeV, with only a
notable exception, the first excited S-wave state of the $bnn$ system.
We cannot say nothing about the P-wave states, but we can expect that
the results from the $V_M$ potential are also close to those of the
$V_Y$ interaction. Thus, it appears that the potential mixing equally
the half perimeter and the centre of mass junction simulates quite well
the genuine Y-shape interaction.

Let us now discuss the quality of the spectra obtained with $V_C$
and $V_G$ and with a renormalized string tension as explained
previously. We performed the calculation with a two-body confining
potential of type $C$ and with a string constant either
$\sigma_r^{(C)}(x)=\sigma\, r_{Y/C}(x)$ (column~(3)) or
$\sigma_R^{(C)}=\sigma\, R_{Y/C}$ (column~(5)). Then we redo the
calculation with a one-body confining potential of type $G$ and
with a string constant either $\sigma_r^{(G)}(x)=\sigma\,
r_{Y/G}(x)$ (column~(4)) or $\sigma_R^{(G)}(x)=\sigma\,
R_{Y/G}(x)$ (column~(6)).

By comparing data from columns (3) and (4) in Table~\ref{tab2}, we
can see that the masses for baryons $nnn$ and $bbb$ are the
practically the same (differences are at the level of the 6th
digit as described in the previous section) and very close to the
value of column (1). This means that, for this kind of symmetrical
systems, the simulation of the potential $V_Y$ by interactions
$V_C$ and $V_G$, based on hyperspherical formalism, gives very
good results. The situation is less favourable for asymmetrical
systems. For $snn$ and $nss$ baryons, the agreement between
columns (3) and (4) is still good, especially for the ground
state. But for very asymmetrical systems such as $bnn$ and $nbb$
baryons, there can exist greater mass differences, up to about 100
MeV. Again, for the ground state, the agreement is still
reasonable. Let us remark that masses in column (4) are always
greater than the corresponding ones in column (3). The values of
these two columns are in reasonable agreement with the reference
results of column (1).

Within the geometrical approach, the results of the two procedures
of simulation are never in perfect agreement, as we can see in
comparing columns (5) and (6), but the values obtained generally
enclose the value of column (1). Let us note that the differences
between a mass in column (5) and the corresponding one in column
(6) are generally of the same order than the gap between values of
columns (3) and (4).

The renormalization of the string tension as suggested by the
geometrical and the hyperspherical formalisms allows to compute
with potentials $V_C$ and $V_G$ binding energies which are close
to the ones obtained with the potential $V_Y$. It gives in any
case much better results than the ones obtained keeping a fixed
value of the string tension.


To be complete, it is worth mentioning that the same conclusions
can be obtained with a relativistic kinematics, although in this
case we cannot compute the reference energies corresponding to the
potential $V_Y$.

\subsection{Particles with different masses}

Up to now, the analysis has been done for at least two identical
particles. General analytical formulas in the case of three different
particles are not available, except $b_G$ and $b_C$ in the
hyperspherical formalism. In this case, we noted that the
tensions are expressed by
\begin{eqnarray}
b_G&=&\frac{32}{15 \pi} \sigma \sum_{i<j} \frac{\beta_{ij}}{\omega_k},
\\
b_C&=&\frac{32}{15 \pi} \sigma \sum_{i<j} \frac{1}{2\alpha_{ij}}.
\end{eqnarray}

A first possibility to treat easily the problem of three different
masses is to use the hypercentral approximation with
$b_M=(b_C+b_G)/2$, with the above expressions for $b_C$ and $b_G$.
This procedure would avoid a double numerical integration to get
$b_Y$, whereas allowing results better than 2\%.

Another possibility relies on the $V_M$ potential proposed in the
previous section. We have verified that this approximation works
well in the case of two or three identical particles. We can
reasonably assume that it will be also good in the case of three
different masses.

\section{Conclusion}
\label{sec5}

In this paper, we studied three approximations of the genuine
three-body confinement: a two-body potential $V_C$ equal to half
perimeter of the triangle formed by the three particles, a
one-body potential $V_G$ with the junction point at the centre of
mass, and a mixing of both $V_M=(V_C+V_G)/2$. Two approaches were
investigated to test the quality of these approximations: a
geometrical one for which the important quantities are the various
distances in the plane of the particles, another one based on the
hyperspherical formalism. Both give very similar and consistent
conclusions. The potential energy $V_G$ overestimates the potential
energy of the genuine junction by about 5\% in most cases, and about
10\% in extreme asymmetrical situation. The confining potential energy
$V_C$ underestimates the potential energy of the genuine junction by
about 8\%. Keeping the same value of the string tension in approximants
can induced a 100 MeV error in the calculated masses, as compared to the
spectra obtained from the Y-shape confinement. In this respect, the
$V_M$ interaction simulates the potential energy of the Y-shape
interaction to better than 2\%.

Thus, using $V_M$ with the same string tension than the genuine junction
gives very good results at the level of the spectra. To obtain similar
quality (sometimes a bit better or a bit worse) for the potentials $V_C$
and $V_G$, it is necessary to renormalize the string tension by a mass
dependent factor that can be analytically computed in the cases of two
and three identical particles. This is very important to simplify the
technical effort.

\section{Acknowledgments}

The authors are grateful to NATO services which give us the possibility
to visit ourselves and greatly facilitates this work, through the NATO
grant PST.CLG.978710. B. Silvestre-Brac and C. Semay (FNRS Research
Associate
position) would like to thank the agreement CNRS/CGRI-FNRS for financial
support.

\clearpage

\begin{table}[htbp]
\caption{Simulation of interaction $V_C$ by interaction $V_G$, and
vice versa, for a nonrelativistic kinematics. Binding energy in
GeV of various baryons as a function of the total orbital angular
momentum $L$, the total angular momentum and parity $J^P$, the
principal quantum number $N$, the type of confinement potential
(Conf.), and the value of the string junction $\sigma$ in GeV$^2$.
The total spin is equal to $J$ and the total isospin is the lowest
one. The mass ratio $x$ is also given for each system. Data
columns are numbered to make the discussion easier. \label{tab1}}
\begin{tabular}{ccccccccccc}
\hline\hline
system & $L$ & $J^P$ & $N$ & Conf. &
$V_C$ & $V_G$ & $V_G$ & $V_G$ & $V_C$ & $V_C$ \\
& & & & $\sigma$ &
$0.2$ & $0.2 r_{C/G}$ & $0.2/R_{G/C}$ &
$0.2$ & $0.2\, r_{G/C}$ & $0.2\, R_{G/C}$ \\
& & & & & (1) & (2) & (3) & (4) & (5) & (6) \\
\hline
$nnn$ & 0 & $1/2^+$ & 1 & &
1.912 & 1.912 & 1.933 & 2.104 & 2.104 & 2.081 \\
($x=1$)& & & 2 & &
2.633 & 2.633 & 2.662 & 2.898 & 2.898 & 2.867 \\
& 1 & $1/2^-$ & 1 & &
2.332 & 2.332 & 2.358 & 2.567 & 2.567 & 2.539 \\

$bbb$ & 0 & $3/2^+$ & 1 & &
0.763 & 0.763 & 0.771 & 0.839 & 0.839 & 0.830 \\
($x=1$)& & & 2 & &
1.050 & 1.050 & 1.062 & 1.156 & 1.156 & 1.143 \\
& 1 & $1/2^-$ & 1 & &
0.930 & 0.930 & 0.940 & 1.024 & 1.024 & 1.013 \\

$snn$ & 0 & $1/2^+$ & 1 & &
1.819 & 1.821 & 1.827 & 2.004 & 2.002 & 1.996 \\
($x=1.667$)& & & 2 & &
2.480 & 2.505 & 2.513 & 2.758 & 2.730 & 2.721 \\
& 1 & $1/2^-$ & 1 & &
2.192 & 2.214 & 2.221 & 2.437 & 2.413 & 2.405 \\

$bnn$ & 0 & $1/2^+$ & 1 & &
1.652 & 1.673 & 1.604 & 1.854 & 1.831 & 1.909 \\
($x=15.76$)& & & 2 & &
2.171 & 2.300 & 2.205 & 2.548 & 2.405 & 2.509 \\
& 1 & $1/2^-$ & 1 & &
1.939 & 2.037 & 1.953 & 2.256 & 2.149 & 2.241 \\

$nss$ & 0 & $1/2^+$ & 1 & &
1.719 & 1.721 & 1.744 & 1.894 & 1.891 & 1.866 \\
($x=0.6$)& & & 2 & &
2.335 & 2.359 & 2.390 & 2.595 & 2.569 & 2.536 \\
& 1 & $1/2^-$ & 1 & &
2.067 & 2.086 & 2.114 & 2.296 & 2.275 & 2.245 \\

$nbb$ & 0 & $1/2^+$ & 1 & &
1.247 & 1.278 & 1.246 & 1.373 & 1.339 & 1.374 \\
($x=0.063$)& & & 2 & &
1.514 & 1.599 & 1.559 & 1.718 & 1.626 & 1.668 \\
& 1 & $1/2^-$ & 1 & &
1.397 & 1.465 & 1.428 & 1.574 & 1.501 & 1.540 \\
\hline\hline
\end{tabular}
\end{table}

\begin{table}[htbp]
\caption{Comparison between genuine potential $V_Y$ with a
constant string tension $\sigma$, the confining potential $V_M$
with the same string tension, and the $V_C$ and $V_G$ confining
interactions with renormalized string tension (see text). Binding
energies are obtained for a nonrelativistic kinematics. Same kind
of data as in Table~\ref{tab1}. \label{tab2}}
\begin{tabular}{ccccccccccc}
\hline\hline system & $L$ & $J^P$ & $N$ & Conf. &
$V_Y$ & $V_M$ & $V_C$ & $V_G$ & $V_C$ & $V_G$ \\
& & & & $\sigma$ & 0.2 & 0.2 & $0.2\, r_{Y/C}$ & $0.2\, r_{Y/G}$ &
$0.2\, R_{Y/C}$ & $0.2/R_{G/Y}$ \\
& & & &  & (1) & (2) & (3) & (4) & (5) & (6)\\
\hline
$nnn$ & 0 & $1/2^+$ & 1 & &
2.032 & 2.009 & 2.036 & 2.036 & 2.020 & 2.040 \\
($x=1$)& & & 2 & &
2.785 & 2.768 & 2.804 & 2.804 & 2.782 & 2.810 \\
& 1 & $1/2^-$ & 1 & &
& 2.451 & 2.483 & 2.483 & 2.464 & 2.488 \\

$bbb$ & 0 & $3/2^+$ & 1 & &
0.810 & 0.801 & 0.812 & 0.812 & 0.806 & 0.814 \\
($x=1$)& & & 2 & &
1.111 & 1.104 & 1.118 & 1.118 & 1.110 & 1.121 \\
& 1 & $1/2^-$ & 1 & &
& 0.978 & 0.990 & 0.990 & 0.983 & 0.993 \\

$snn$ & 0 & $1/2^+$ & 1 & &
1.933 & 1.913 & 1.935 & 1.937 & 1.921 & 1.928 \\
($x=1.667$)& & & 2 & &
2.626 & 2.625 & 2.639 & 2.666 & 2.620 & 2.653 \\
& 1 & $1/2^-$ & 1 & &
& 2.317 & 2.332 & 2.356 & 2.316 & 2.345 \\

$bnn$ & 0 & $1/2^+$ & 1 & &
1.752 & 1.760 & 1.747 & 1.769 & 1.745 & 1.695 \\
($x=15.76$)& & & 2 & &
2.310 & 2.390 & 2.295 & 2.432 & 2.293 & 2.330 \\
& 1 & $1/2^-$ & 1 & &
& 2.110 & 2.050 & 2.153 & 2.049 & 2.063 \\

$nss$ & 0 & $1/2^+$ & 1 & &
1.826 & 1.807 & 1.828 & 1.831 & 1.815 & 1.841 \\
($x=0.6$)& & & 2 & &
2.474 & 2.468 & 2.484 & 2.509 & 2.467 & 2.523 \\
& 1 & $1/2^-$ & 1 & &
& 2.183 & 2.199 & 2.219 & 2.184 & 2.231 \\

$nbb$ & 0 & $1/2^+$ & 1 & &
1.313 & 1.312 & 1.296 & 1.329 & 1.317 & 1.316 \\
($x=0.063$)& & & 2 & &
1.645 & 1.618 & 1.574 & 1.662 & 1.599 & 1.647 \\
& 1 & $1/2^-$ & 1 & &
& 1.488 & 1.453 & 1.523 & 1.476 & 1.509 \\
\hline\hline
\end{tabular}
\end{table}

\clearpage



\begin{figure}
\caption{Ratio $R_{G/C}(x)$ from the geometrical treatment (see
formula~(\ref{pgc})) and ratio $r_{G/C}(x)=b_G(x)/b_C(x)$ from the
hyperspherical treatment (see formulas~(\ref{bg}) and (\ref{bc})), as a
function of the mass ratio $x$.}
\label{fig3}
\end{figure}


\begin{figure}
\caption{Ratio $R_{G/Y}(x)$ from the geometrical treatment (see
formula~(\ref{rgy})) and ratio $r_{G/Y}(x)=b_G(x)/b_Y(x)$ from the
hyperspherical treatment (see formulas~(\ref{bg}) and (\ref{by})), as a
function of the mass ratio $x$.}
\label{fig5}
\end{figure}

\begin{figure}
\caption{Ratio $R_{M/Y}(x)$ from the geometrical treatment (see
formula~(\ref{rmy})) and ratio $r_{M/Y}(x)$ from the
hyperspherical treatment (see formula~(\ref{rmy2})), as a function
of the mass ratio $x$.} \label{figsup}
\end{figure}

\begin{figure}
\caption{Integration domain of quantity $L(R,\theta,\chi)$ as a function
of angles $\theta$ and $\chi$ (see Sect.~\ref{hyper}).}
\label{fig6}
\end{figure}

\begin{figure}
\caption{Ratio $R_{Y/C} \approx 1.086$ from the geometrical treatment
(see formula~(\ref{pe})) and ratio $r_{Y/C}(x)=b_Y(x)/b_C(x)$ from the
hyperspherical treatment (see formulas~(\ref{by}) and (\ref{bc})), as a
function of the mass ratio $x$.}
\label{fig7}
\end{figure}

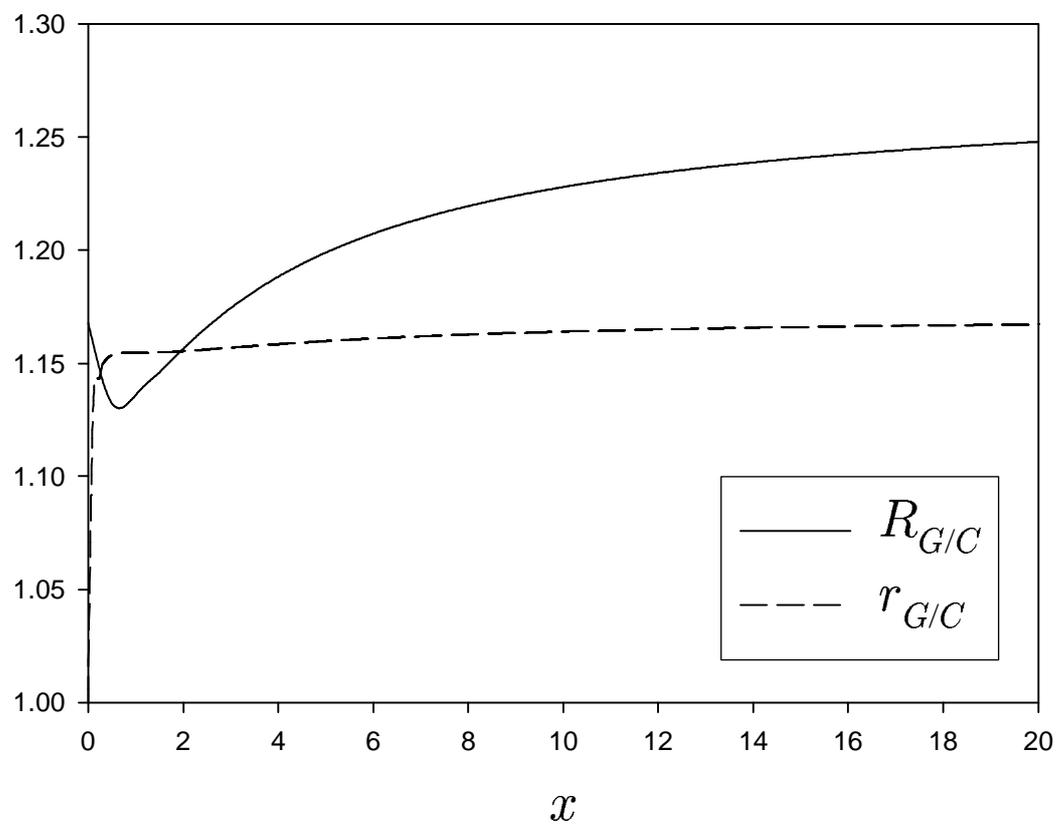
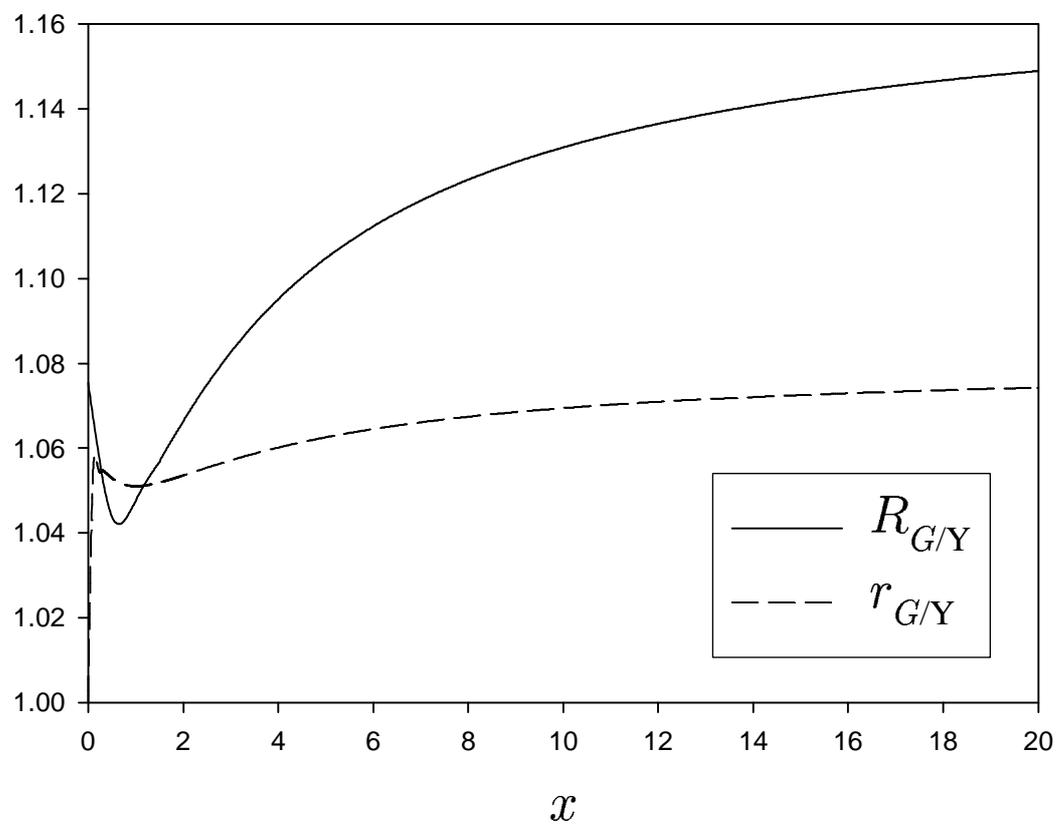
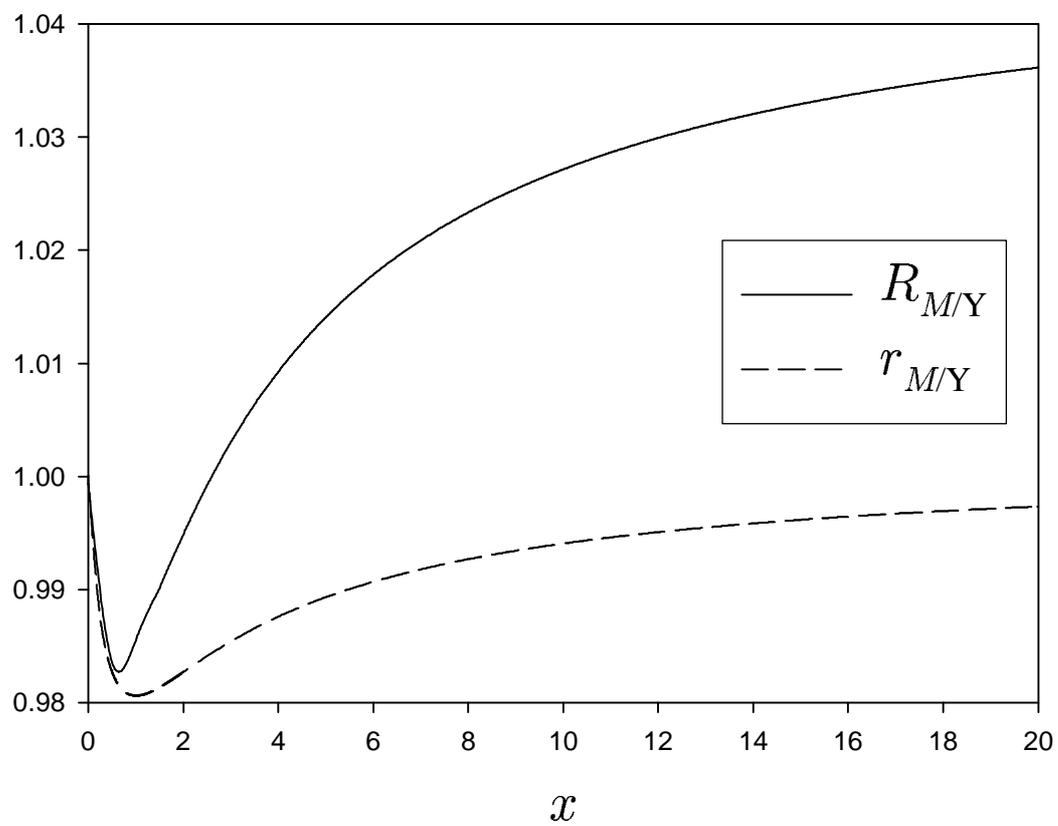
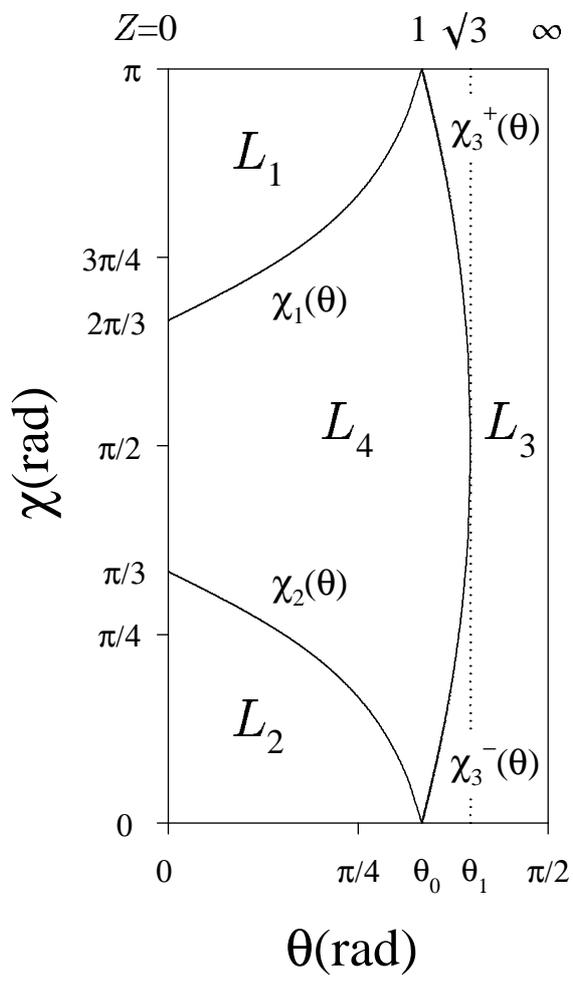
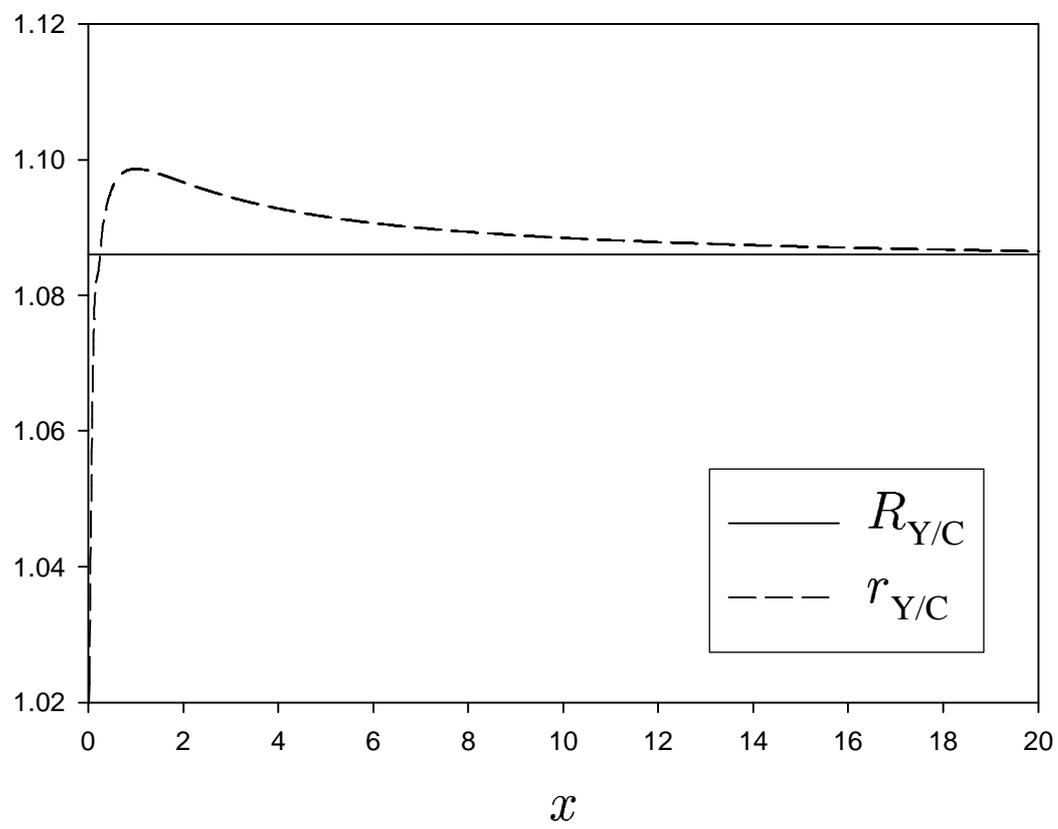
\end{document}